\documentclass[prl,showpacs,twocolumn,showkeys]{revtex4-1}
\usepackage{multirow}
\usepackage{amsmath}
\usepackage{amssymb}
\usepackage{graphics}
\usepackage{graphicx}
\usepackage{array}
\usepackage{color}
\usepackage[hypertex, colorlinks, colorlinks=false, linkcolor=blue, citecolor=green, filecolor=black, urlcolor=cyan]{hyperref}

\newcommand{\bibs}{c:/users/rlyte/dropbox/AllSharedFolders/References/BibFile}

\begin{document}

\title{Exact fundamental limits of the first and second hyperpolarizabilities}

\author{Rick Lytel}
\email[]{rlytel@wsu.edu}
\author{Sean Mossman}
\author{Ethan Crowell}
\author{Mark G. Kuzyk}
\affiliation{Department of Physics and Astronomy, Washington State University, Pullman, Washington  99164-2814}

\date{\today}

\begin{abstract}
Nonlinear optical interactions of light with materials originate in the microscopic response of the molecular constituents to excitation by an optical field, and are expressed by the first ($\beta$) and second ($\gamma$) hyperpolarizabilities. Upper bounds to these quantities were derived seventeen years ago using approximate, truncated state models that violated completeness and unitarity, and far exceed those achieved by potential optimization of analytical systems. This letter determines the fundamental limits of the first and second hyperpolarizability tensors using Monte-Carlo sampling of energy spectra and transition moments constrained by the diagonal Thomas-Reiche-Kuhn (TRK) sum rules and filtered by the off-diagonal TRK sum rules. The upper bounds of $\beta$ and $\gamma$ are determined from these quantities by applying error-refined extrapolation to perfect compliance to the sum rules.  The method yields the largest diagonal component of the hyperpolarizabilities for an arbitrary number of interacting electrons in any number of dimensions. The new method provides design insight to the synthetic chemist and nanophysicist for approaching the limits. This analysis also reveals that the special cases which lead to divergent nonlinearities in the many-state catastrophe are not physically realizable.
\end{abstract}


\maketitle

\textit{Introduction}--Nonlinear optics is the study of quantum systems with polarizations that are nonlinear functions of external electromagnetic fields.  This letter solves the problem of determining the exact fundamental limits of nonlinear optics by delineating the first procedure for computing the first and second hyperpolarizabilities consistent with on- and off-diagonal quantum mechanical sum rules.  In the process, we show that prior predictions of the limits using truncated sum rules are too high by nearly $30\%$ for $\beta$\cite{kuzyk00.01,kuzyk03.01} and $40\%$ for $\gamma$\cite{kuzyk03.01}, that predictions of the many-state catastrophe\cite{shafe13.01} are spurious, and that predictions of the scaling of the hyperpolarizabilities with the strength of the ground-to-excited state transition moment are modified in a way that will direct molecular synthesists to make new design choices.

The nonlinear optical response of a material is generated by the collective response of the basic elements comprising it.  This letter concerns the maximum values of the nonlinear optical response of a molecular-scale structure, not a material.  The nonlinear optics of an elemental structure is measured by the effect it has on the molecular polarization vector when perturbatively excited by an electric field $\mathcal{E}_{i}$, with $i=x,y,z$:
\begin{equation}\label{polVec}
p_{i}=\mu_{i}+\alpha_{ij}\mathcal{E}_{j}+\beta_{ijk}\mathcal{E}_{j}\mathcal{E}_{k}+\gamma_{ijkl}\mathcal{E}_{j}\mathcal{E}_{k}\mathcal{E}_{l}
\end{equation}
where $\mu_{i}$ is $i^\text{th}$ component of the ground state dipole moment vector, $\alpha_{ij}$ is the linear polarizability tensor, $\beta_{ijk}$ is the first hyperpolarizability tensor, $\gamma_{ijkl}$ is the second hyperpolarizability tensor, and repeated indices are summed.

The computation of the first and second hyperpolarizabilities can be accomplished in perturbation theory using a sum over states\cite{ward65.02} (SOS), Dalgarno-Lewis perturbation theory\cite{dalga55.01,mossm16.01,lytel16.01}, the method of finite fields\cite{zhou07.02}, and others.  Each method requires state and spectral information from a Hamiltonian.  In this letter, we focus on the largest off-resonant, diagonal tensor component of the first hyperpolarizability for which the SOS expression is given by
\begin{equation}\label{beta}
\beta_\text{SOS}\equiv\beta_\text{xxx} = 3e^3 {\sum_{n,m}}' \frac{x_{0n} \bar{x}_{nm} x_{m0}}{E_{n0} E_{m0}},
\end{equation}
where the prime on the sum indicates omission of the ground state, $e$ is the electron charge, $x_{nm}=\langle n|x|m\rangle$ is the many-body transition moment, $\bar{x}_{nm}=x_{nm}-\delta_{nm}x_{00}$, and $E_{nm}=E_{n}-E_{m}$ is the difference between energy eigenvalues. The SOS expression is an exact solution for $\beta$.

Completeness of the states leads directly to the Thomas-Reiche-Kuhn sum rules\cite{thom25.01,reich25.01,kuhn25.01}, an infinite set of equations relating the transition moments and spectra
\begin{equation}\label{TRKsumrule}
S_{nm} = \sum_{p=0}^{\infty} [E_{pn}+E_{pm}] x_{np}x_{pm}=\frac{N_e\hbar^2}{m}\delta_{nm},
\end{equation}
where $N_e$ is the number of electrons and the spatial operator $x$ represents any Cartesian direction.

The fundamental limits can be calculated in principle by determining the maximum value of the SOS expression given by Eq. \ref{beta} constrained by the sum rules, Eq. \ref{TRKsumrule}, which reduces the number of free parameters to make it possible to find an extremum. However, after years of effort, nobody has succeeded in implementing an analytical method that yields an algebraic expression.

\textit{Three-level model limits}--The first effort to determine an upper bound used a three-level SOS model (TLM) for $\beta$ with truncated sum rules to estimate a limit $\beta_\text{max}$\cite{kuzyk00.01}.  The fact that all experimental data at that time fell at least a factor of 30 below the limit supported the results,\cite{kuzyk03.02} but also raised questions about the gap.  Similar calculations determined the limits for $\gamma$\cite{kuzyk00.02}. The usefulness of the theory was in evidence when it was soon after applied to analyze hyper-Rayleigh scattering experiments\cite{clays01.01}.  Dividing $\beta$ by this maximum yields an intrinsic first hyperpolarizability $\beta_\text{int}\leq 1$, which has been found to be a useful choice of units for comparing a vast array of quantum systems because it is scale-free and effectively measures the efficient use of the electrons by a molecular structure,\cite{kuzyk13.01}.  For $\gamma_\text{int}$, the limits are $-0.25\leq\gamma_\text{int}\leq 1$.  We use the intrinsic values in the remainder of this paper.

The truncation of the SOS expression to three energy eigenstates assumes that only these states contribute significantly when $\beta_\text{int}$ is near its maximum value, a manifestation of the three-level ansatz (TLA)\cite{kuzyk00.01} and an assertion of convergence.  The full sum rules remain satisfied because all other states exist, but have a negligible contribution to the hyperpolarizabilities\cite{kuzyk06.03}.  On the other hand, truncation of the sum rules to three states violates completeness and necessarily yields transition moments which cannot result from a mechanical Hamiltonian.  Truncation of the sum rules is tantamount to assuming no other states exist, a physical impossibility.  Hence, the limit of unity for $\beta_\text{int}$ -- which was derived by truncating the sum rules -- may not be an accurate estimate of the true fundamental limit\cite{lytel16.02}.

Optimizations of parameterized potentials\cite{zhou07.02,watki09.01,watki11.01,ather12.01,burke13.01} and computations of model Hamiltonians\cite{lytel13.04} corroborate this conclusion.  Each produces a maximum value of $\beta_\text{int}$ of about $0.71$, nearly $30\%$ lower than that predicted by the three level model. The apparent limits for $\gamma_\text{int}$ are $40\%$ of those predicted by the three level model.  This is the so-called \emph{limit gap} between physical systems--the \emph{Hamiltonian Limits}--and the original limits, and until now, its origin has remained elusive.

\textit{Many-state Monte-Carlo limits}--A statistical approach \emph{with many more than three states} was developed to explore a large set of spectra and moments, and generated results supporting the limit of unity predicted by the three-level model\cite{kuzyk08.01}.  This Monte Carlo (MC) method used a dipole-free sum over states\cite{kuzyk05.02,champ05.01} (DFSOS) expression
\begin{equation}\label{betaDF}
\beta_\text{DFSOS} = 3e^3 {\sum_{n\neq m}}' \frac{x_{0n} x_{nm} x_{m0}} {E_{n0} E_{m0}}\Bigg[1-\frac{E_{m0}(2E_{m0}-E_{n0})}{E_{n0}^2}\Bigg],
\end{equation}
which eliminates the diagonal transition moments $x_{nn}$.  The DFSOS model was applied to the study of the first hyperpolarizability of push-pull $\pi$-conjugated systems and compared favorably to the standard sum over states\cite{champ06.01}, where they agree well in the static limit studied here.

The method generates a set of random energy eigenvalues and off-diagonal transition moments by enforcing the diagonal TRK sum rules
\begin{equation}\label{TRKdiag}
S_{nn}=2\sum_{p=0}^{N_{s}}E_{pn}|x_{np}|^2=\frac{N_e\hbar^2}{m}
\end{equation}
for $n=0,1,2...N_{s}-1$. Here, $N_{s}$ is the number of excited states used in the model. [In this letter, all results use $N_{s}=21$.  This choice is much larger than the number of states (about ten, in our experience) above which the results remain unchanged.]  Note that the diagonal sum rules only involve the energy spectrum and the off-diagonal transition moments.  Through a clever manipulation of Eqs. \ref{TRKdiag}, the off-diagonal moments are generated at random and used in the DFSOS expression, Eq. \ref{betaDF}, to generate a large ensemble of $\beta$ values.  Since the DFSOS expression does not require the diagonal moments, the off-diagonal sum rules are ignored, and the diagonal moments are left unspecified. While the DFSOS expression agrees perfectly with the SOS expression within the three-state model and in the infinite state limit, we will show how neglecting the diagonal transition moments can lead to internal inconsistencies within the off-diagonal sum rules and correcting this approximation produces results which are consistent with the Hamiltonian upper limits of $|\beta_\text{int}| \leq 0.71$ and $-0.15\leq\gamma_\text{int}\leq 0.6$.

\textit{Filtered Monte Carlo limits}--The inconsistency is revealed when one tries to identify the diagonal transition elements which were omitted above. The above algorithm results in a set of off-diagonal transition moments $x_{n\neq m}$ and energies $E_{n0}$ which identically satisfy the diagonal sum rules, Eqs. \ref{TRKdiag}. One set of off-diagonal sum rules, $S_{n0}$ with $n>0$
\begin{align}
    S_{n0} = \sum_{p\neq n}^{N_s}\left(E_{pn}+E_{p0}\right)x_{np}x_{p0}+E_{n0}x_{nn}x_{n0} =0,
    \label{SRn0}
\end{align}
for example, fully defines the previously unspecified diagonal transition elements by
\begin{align}
     x_{nn}=-\frac{\sum_{p\neq n}^{N_s}\left(E_{pn}+E_{p0}\right)x_{np}x_{p0}}{E_{n0}x_{n0}},
     \label{xnn}
\end{align}
where we recognize that $x_{00}$ is a free parameter which we can always take to be zero.

Any other column of off-diagonal sum rules provides equally legitimate determinations of the diagonal transition elements, but in general, these results are inconsistent with one another. It must be possible to generate self-consistent sum rules as this is what we expect of systems generated from physical Hamiltonians.

Thus, the previous MC algorithm can be modified as follows.  First, a set of random spectra are created, and then the random off-diagonal moments $x_{nm}$ are computed using the diagonal sum rules as in previous works\cite{kuzyk08.01}. The diagonal moments are then selected by using Eqs. \ref{SRn0}, picking out a particular set of diagonal transition elements. Out of millions of MC runs, it is probable that some finite number of these selections may lead to off-diagonal sum rules, $S_{nm}$ for $n\neq m$, that are self-consistently satisfied to some tolerance.  To find these, we compute the standard deviation $\sigma_{N}$ defined as
\begin{equation}\label{sigmaN}
\sigma_{N}=\frac{2}{N(N+1)}\sqrt{\sum_{n<m}^{N}S_{nm}^2}
\end{equation}
and find the MC runs for which $\sigma_{N}<\epsilon$, where $\epsilon$ is a tolerance number close to zero.  For these runs, with the $S_{n0}$ satisfied by Eq. \ref{SRn0}, the next lowest $N(N-1)/2$ off-diagonal sum rules computed in the $N_{s}$ state model are also nearly satisfied.  The integer $N$ is simply the largest value of $m$ for which the $S_{nm}, n<m$ are nearly satisfied.  Solutions meeting this criterion are deemed the filtered (and nearly valid) SOS solutions.  The modified algorithm generates millions of values for $\beta_\text{int}$ and $\gamma_\text{int}$, but the filtering process picks out those values whose transition elements and energies satisfy the off-diagonal sum rules to the tolerance selected.  By varying the tolerance and the number of sum rules enforced, one can determine the limiting values to which the filtered values of $\beta_\text{int}$ and $\gamma_\text{int}$ asymptote.  This is the self-consistent MC algorithm.

\begin{figure}\centering
\includegraphics[width=3.4in]{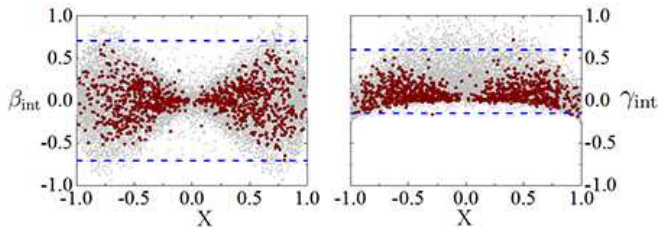}
\vspace{-0.2cm}
\caption{Monte Carlo simulation of the first and second intrinsic hyperpolarizabilities with one million instances (light grey points), and the modified version where the off-diagonal sum rules are enforced to a limit $\sigma_{6}\leq 0.12$ (dark red points). The filtered values constitute approximately $1\%$ of the unfiltered values. The dashed lines indicate the Hamiltonian Limits.}\label{fig:betaGammaRandom}
\end{figure}

Fig \ref{fig:betaGammaRandom} shows the result of this modification in a scatter plot showing $\beta_\text{int}$ and $\gamma_\text{int}$ plotted against the $X$ parameter, $X\equiv x_{01}/x_{01}^\text{max}$, the independent TLM parameter measuring the fraction of oscillator strength captured in the ground-to-first excited state transition and a convenient independent parameter for the Monte Carlo calculations\cite{kuzyk08.01,shafe11.01}.  In the figure, one million MC instances were run, and a total of about one thousand were found for which the standard deviation of the sum rule $\sigma_{5}<\epsilon$, with $\epsilon=0.12$ as the tolerance for the off-diagonal sum rules. Thus, approximately one in a thousand MC runs satisfy the off-diagonal sum rules to this tolerance, yielding a set of values of $\beta_\text{int}$ that are indicated as dark red points in Fig \ref{fig:betaGammaRandom}.  The light points are the one million unfiltered values calculated using the DFSOS expression corresponding to the MC algorithm of previous works.  The dark red points fall between the dashed lines for both $\beta_\text{int}$ $(\beta_\text{int}\leq 0.7)$ and $\gamma_\text{int}$ $(-0.15 \leq \gamma_\text{int}\leq 0.6)$, suggesting that when the off-diagonal sum rules are obeyed, the fundamental limit is lower than previously calculated.

\begin{figure}\centering
\includegraphics[width=3.4in]{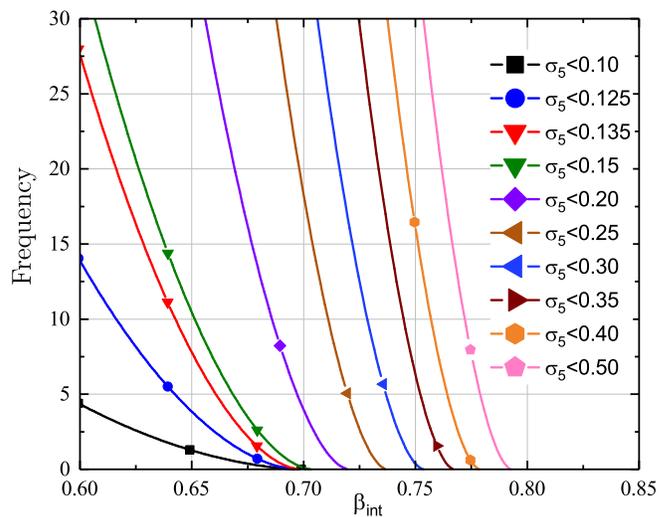}
\vspace{-0.2cm}
\caption{Cycloid fits to the frequency of occurrence of $\beta_\text{int}$ for ten values of $\sigma_{5}\leq \epsilon$, which illustrates how the largest value of $\beta_\text{int}$ decreases as the tolerance on the off-diagonal sum rules $\epsilon$ is decreased.}\label{fig:30M_TEST_6_state_freqVsBeta}
\end{figure}

\begin{figure}\centering
\includegraphics[width=3.4in]{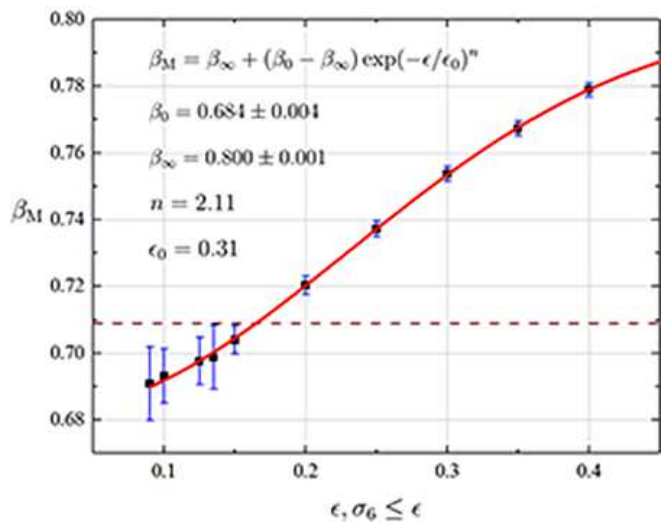}
\vspace{-0.2cm}
\caption{The largest intrinsic hyperpolarizability  $\beta_\text{M}$ -- as determined from the cycloid fits as shown in Fig \ref{fig:30M_TEST_6_state_freqVsBeta}, as a function of the off-diagonal sum rule tolerance, $\epsilon$ (points).  The solid curve is a fit to the function shown in the inset.   The empirically-determined Hamiltonian limit is indicated with a dashed line.}\label{fig:30M_TEST_6_state_betaVsEpsilon}
\end{figure}

A more objective estimate of the new limit requires a two-step process.  First, the distribution is fit to a function with a sharp cutoff, where the cut-off indicates the limit.  We use the cycloid function $f=f_{0}[1-(\beta_\text{int}/\beta_\text{M})^{1/n}]^n$, where $f$ is the frequency of occurrence and $\beta_\text{M}$ is the apparent limit under a given sum rule tolerance.  Fig \ref{fig:30M_TEST_6_state_freqVsBeta} shows such cycloid fits as a function of the sum rule tolerance associated with the first six states. The cutoff is seen to converge near $\beta_\text{int}\approx 0.7$ as the constraint is tightened.  Next, the cutoff hyperpolarizability, as determined from each cycloid fit, is plotted as a function of sum rule tolerance $\epsilon$, as shown in Fig \ref{fig:30M_TEST_6_state_betaVsEpsilon}.  The points are generated by running thirty million MC instances with the additional pruning algorithm.  This data in turn is fit to the function $\beta_{\text{M}} = \beta_\infty + \left( \beta_0 - \beta_\infty \right) \exp(-\epsilon / \epsilon_0)^n$, which captures the asymptote to the new upper bound in the limit of zero tolerance.   The intercept $\beta_0$ obtained from the fit thus gives an estimate of the true hyperpolarizability limit.  As the tolerance is made tighter, the filtered MC results become much more rare, creating greater uncertainty, hence the larger error bars near the $\epsilon = 0$ limit.  The results suggest that the actual limiting value is near $\beta_\text{M} \approx 0.7$, the Hamiltonian limit.

\begin{figure*}\centering
    \includegraphics[width=3.3in]{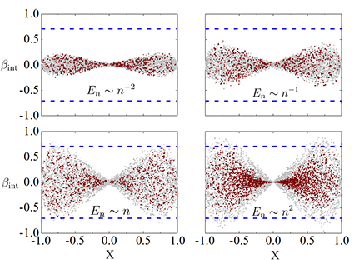}
    \includegraphics[width=3.3in]{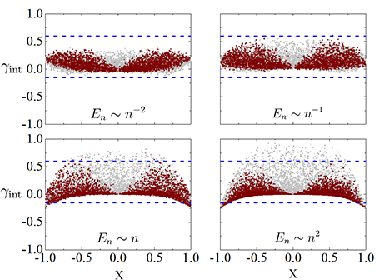}
    \vspace{-0.2cm}
    \caption{Monte Carlo constrained by $S_{nn}$ only (light grey) and filtered (dark red) values of $\beta_\text{int}$ (left) and $\gamma_\text{int}$ (right) assuming different energy spectra scaling, illustrating the convergence of the limits for the hyperpolarizabilities to the Hamiltonian Limits, indicated by dashed lines.}
    \label{fig:QuadEnergies}
    \vspace{-0.2cm}
\end{figure*}

\textit{New results}--The computational method delineated in this letter reveals first and foremost that the \emph{limit gap} between the Hamiltonian Limits and the fundamental limits previously calculated, argued to originate from the fact that the sum rules apply to a more general class of Hamiltonians\cite{kuzyk13.01}, does not exist. This implies that conventional molecular designs should be able to achieve the true fundamental limits of nonlinear optics.

Following the first MC paper\cite{kuzyk08.01}, another was published to study the effect of the scaling character of the energy spectrum on the limiting values of $\beta_\text{int}$ and $\gamma_\text{int}$\cite{shafe13.01}. Demonstrating the degree to which the diagonal sum rules capture the majority of the physics, the results in that paper are qualitatively correct in that the largest values are obtained for spectra scaling linearly or faster with eigenstate number. But the quantitative limits for each spectra are the Hamiltonian Limits, not the three-level model limits, as shown in Fig \ref{fig:QuadEnergies}. Most important--and a key result of this letter-- is a striking quantitative difference in the shape of the distributions for $\gamma$ such that the regions of strongest response no longer cluster around $X=0$ but rather around a range of $X$ similar to the maximum value for $\beta$. Our work reveals that the origin of this fundamental feature is that when $x_{01}=0$, the most important contributions to $\gamma$ from the sum over states vanish, a fact that was not noticed in the original work, which ignored the dipole moments and generated spurious results.  The best possible $\beta$ systems fall within a similar parameter space to the best possible $\gamma$ systems. This result suggests molecules that yield large $\beta$ could also yield large $\gamma$, barring symmetry constraints.  The result also motivates future research on the filtered MC algorithm for exploring symmetries in nonlinear optics beyond those that are determined by analytic methods.

In 2013, researchers applied the unfiltered MC algorithm finding that a specific energy spectrum, where the first excited state is many-fold degenerate and the second state is taken to be at a very high energy, could lead to an arbitrarily high hyperpolarizability in the limit of an infinite number of degenerate states\cite{shafe13.01}. The filtering procedure reported in this work, when applied to such energy spectra, effectively eliminates them, indicating that a fundamental quantum mechanical constraint disallows such spectra. This conclusion is consistent with attempts to solve the inverse problem for such spectra\cite{dawso16.02}. Our work thus resolves the many-state catastrophe by suggesting that it is disallowed.

The modified Monte Carlo approach described in this letter generates sets of spectra and transition moments that allow the first and second hyperpolarizabilities to approach the Hamiltonian Limits, $|\beta|\leq 0.71$ and $-0.15\leq\gamma\leq 0.6$.  As noted and referenced above, the best potential models achieve these limits.  Quasi-one dimensional, many electron structures with a linear chain and a side group or prong will generate \emph{phase disruption} among the lowest energy states near the Fermi level, and this phenomena leads to large intrinsic response approaching the limits for both $\beta$ and $\gamma$\cite{lytel15.02}.  Molecules with conjugated chains and variable spacers\cite{stefk13.01} meet these criteria, provided that an appropriate atom providing complete conjugation is placed at the intersection of the main chain and side groups. Nanostructures consisting of short metal rods with a side prong also meet these criteria and are of interest for future exploration.  Finally, hybrid materials\cite{sulli16.02} point to design paradigms which could achieve the exact limits.  These systems have spectra that scale linearly or faster with eigenstate number.  Spectra typical of Coulomb forces within molecular systems scale as an inverse power of the eigenstate number and fall far short of the limits. It is apparent from Fig. \ref{fig:QuadEnergies} that the proper spectra scaling is a necessary condition to reach the limits.

\textit{Conclusion}--In summary, by employing a self-consistent Monte Carlo algorithm, we have the first convincing evidence that the three-level model limits overestimate the actual fundamental limits, corroborating the so-called Hamiltonian Limits and dispelling concern that potential optimizations have simply been stuck on an uncanny local maximum. Enforcing a physical set of off-diagonal sum rules yields accurate estimates of the true fundamental limits on the hyperpolarizabilities.  Furthermore, we have resolved the paradox of the many-state catastrophe by showing that such systems disobey the off-diagonal sum rules, removing any loopholes for exceeding the limits and bringing optimal materials within reach of standard design paradigms.  It is of particular interest that the diagonal sum rules alone are enough to establish fundamental limits to within 30\%, in agreement with the TLM limit, and that a handful of off-diagonal sum rules bring the limit into agreement with potential and topological optimization results.

This work provides a self-consistent sampling broadly applicable to all many-electron systems in three dimensions and demonstrates a powerful tool for exploring nonlinear coupled equations in quantum optics. The open question of how to derive the proper fundamental limits from first principles remains. This  work indicates that the difference between the three-level limits and the true fundamental limits may rest on the interplay between the three-level ansatz and the minimum complexity required for sufficiently consistent sum rules.

\begin{acknowledgments}
We acknowledge NSF (ENG) (ECCS-1128076) and the Meyer Distinguished Professorship of the Sciences.
\end{acknowledgments}

\bibliography{\bibs}

\begin{thebibliography}{31}%
\makeatletter
\providecommand \@ifxundefined [1]{%
 \@ifx{#1\undefined}
}%
\providecommand \@ifnum [1]{%
 \ifnum #1\expandafter \@firstoftwo
 \else \expandafter \@secondoftwo
 \fi
}%
\providecommand \@ifx [1]{%
 \ifx #1\expandafter \@firstoftwo
 \else \expandafter \@secondoftwo
 \fi
}%
\providecommand \natexlab [1]{#1}%
\providecommand \enquote  [1]{``#1''}%
\providecommand \bibnamefont  [1]{#1}%
\providecommand \bibfnamefont [1]{#1}%
\providecommand \citenamefont [1]{#1}%
\providecommand \href@noop [0]{\@secondoftwo}%
\providecommand \href [0]{\begingroup \@sanitize@url \@href}%
\providecommand \@href[1]{\@@startlink{#1}\@@href}%
\providecommand \@@href[1]{\endgroup#1\@@endlink}%
\providecommand \@sanitize@url [0]{\catcode `\\12\catcode `\$12\catcode
  `\&12\catcode `\#12\catcode `\^12\catcode `\_12\catcode `\%12\relax}%
\providecommand \@@startlink[1]{}%
\providecommand \@@endlink[0]{}%
\providecommand \url  [0]{\begingroup\@sanitize@url \@url }%
\providecommand \@url [1]{\endgroup\@href {#1}{\urlprefix }}%
\providecommand \urlprefix  [0]{URL }%
\providecommand \Eprint [0]{\href }%
\providecommand \doibase [0]{http://dx.doi.org/}%
\providecommand \selectlanguage [0]{\@gobble}%
\providecommand \bibinfo  [0]{\@secondoftwo}%
\providecommand \bibfield  [0]{\@secondoftwo}%
\providecommand \translation [1]{[#1]}%
\providecommand \BibitemOpen [0]{}%
\providecommand \bibitemStop [0]{}%
\providecommand \bibitemNoStop [0]{.\EOS\space}%
\providecommand \EOS [0]{\spacefactor3000\relax}%
\providecommand \BibitemShut  [1]{\csname bibitem#1\endcsname}%
\let\auto@bib@innerbib\@empty
\bibitem [{\citenamefont {Kuzyk}(2000{\natexlab{a}})}]{kuzyk00.01}%
  \BibitemOpen
  \bibfield  {author} {\bibinfo {author} {\bibfnamefont {M.~G.}\ \bibnamefont
  {Kuzyk}},\ }\href@noop {} {\bibfield  {journal} {\bibinfo  {journal} {Phys.
  Rev. Lett.}\ }\textbf {\bibinfo {volume} {85}},\ \bibinfo {pages} {1218}
  (\bibinfo {year} {2000}{\natexlab{a}})}\BibitemShut {NoStop}%
\bibitem [{\citenamefont {Kuzyk}(2003{\natexlab{a}})}]{kuzyk03.01}%
  \BibitemOpen
  \bibfield  {author} {\bibinfo {author} {\bibfnamefont {M.~G.}\ \bibnamefont
  {Kuzyk}},\ }\href@noop {} {\bibfield  {journal} {\bibinfo  {journal} {Opt.
  Lett.}\ }\textbf {\bibinfo {volume} {28}},\ \bibinfo {pages} {135} (\bibinfo
  {year} {2003}{\natexlab{a}})}\BibitemShut {NoStop}%
\bibitem [{\citenamefont {Shafei}\ and\ \citenamefont
  {Kuzyk}(2013)}]{shafe13.01}%
  \BibitemOpen
  \bibfield  {author} {\bibinfo {author} {\bibfnamefont {S.}~\bibnamefont
  {Shafei}}\ and\ \bibinfo {author} {\bibfnamefont {M.~G.}\ \bibnamefont
  {Kuzyk}},\ }\href {\doibase 10.1103/PhysRevA.88.023863} {\bibfield  {journal}
  {\bibinfo  {journal} {Phys. Rev. A}\ }\textbf {\bibinfo {volume} {88}},\
  \bibinfo {pages} {023863} (\bibinfo {year} {2013})}\BibitemShut {NoStop}%
\bibitem [{\citenamefont {Ward}(1965)}]{ward65.02}%
  \BibitemOpen
  \bibfield  {author} {\bibinfo {author} {\bibfnamefont {J.~F.}\ \bibnamefont
  {Ward}},\ }\href@noop {} {\bibfield  {journal} {\bibinfo  {journal} {Reviews
  of Modern Physics}\ }\textbf {\bibinfo {volume} {37}},\ \bibinfo {pages} {1}
  (\bibinfo {year} {1965})}\BibitemShut {NoStop}%
\bibitem [{\citenamefont {Dalgarno}\ and\ \citenamefont
  {Lewis}(1955)}]{dalga55.01}%
  \BibitemOpen
  \bibfield  {author} {\bibinfo {author} {\bibfnamefont {A.}~\bibnamefont
  {Dalgarno}}\ and\ \bibinfo {author} {\bibfnamefont {J.~T.}\ \bibnamefont
  {Lewis}},\ }\href@noop {} {\bibfield  {journal} {\bibinfo  {journal} {Proc.
  R. Soc. London Ser. A}\ }\textbf {\bibinfo {volume} {233}},\ \bibinfo {pages}
  {70} (\bibinfo {year} {1955})}\BibitemShut {NoStop}%
\bibitem [{\citenamefont {Mossman}\ \emph {et~al.}(2016)\citenamefont
  {Mossman}, \citenamefont {Lytel},\ and\ \citenamefont {Kuzyk}}]{mossm16.01}%
  \BibitemOpen
  \bibfield  {author} {\bibinfo {author} {\bibfnamefont {S.}~\bibnamefont
  {Mossman}}, \bibinfo {author} {\bibfnamefont {R.}~\bibnamefont {Lytel}}, \
  and\ \bibinfo {author} {\bibfnamefont {M.~G.}\ \bibnamefont {Kuzyk}},\
  }\href@noop {} {\bibfield  {journal} {\bibinfo  {journal} {J. Opt. Soc. Am.
  B}\ }\textbf {\bibinfo {volume} {33}},\ \bibinfo {pages} {E31} (\bibinfo
  {year} {2016})}\BibitemShut {NoStop}%
\bibitem [{\citenamefont {Lytel}\ \emph {et~al.}(2016)\citenamefont {Lytel},
  \citenamefont {Mossman},\ and\ \citenamefont {Kuzyk}}]{lytel16.01}%
  \BibitemOpen
  \bibfield  {author} {\bibinfo {author} {\bibfnamefont {R.}~\bibnamefont
  {Lytel}}, \bibinfo {author} {\bibfnamefont {S.~M.}\ \bibnamefont {Mossman}},
  \ and\ \bibinfo {author} {\bibfnamefont {M.~G.}\ \bibnamefont {Kuzyk}},\
  }\href@noop {} {\bibfield  {journal} {\bibinfo  {journal} {J. Opt. Soc. Am.
  B}\ }\textbf {\bibinfo {volume} {33}},\ \bibinfo {pages} {E14} (\bibinfo
  {year} {2016})}\BibitemShut {NoStop}%
\bibitem [{\citenamefont {Zhou}\ \emph {et~al.}(2007)\citenamefont {Zhou},
  \citenamefont {Szafruga}, \citenamefont {Watkins},\ and\ \citenamefont
  {Kuzyk}}]{zhou07.02}%
  \BibitemOpen
  \bibfield  {author} {\bibinfo {author} {\bibfnamefont {J.}~\bibnamefont
  {Zhou}}, \bibinfo {author} {\bibfnamefont {U.~B.}\ \bibnamefont {Szafruga}},
  \bibinfo {author} {\bibfnamefont {D.~S.}\ \bibnamefont {Watkins}}, \ and\
  \bibinfo {author} {\bibfnamefont {M.~G.}\ \bibnamefont {Kuzyk}},\ }\href@noop
  {} {\bibfield  {journal} {\bibinfo  {journal} {Phys. Rev. A}\ }\textbf
  {\bibinfo {volume} {76}},\ \bibinfo {pages} {053831} (\bibinfo {year}
  {2007})}\BibitemShut {NoStop}%
\bibitem [{\citenamefont {Thomas}(1925)}]{thom25.01}%
  \BibitemOpen
  \bibfield  {author} {\bibinfo {author} {\bibfnamefont {W.}~\bibnamefont
  {Thomas}},\ }\href@noop {} {\bibfield  {journal} {\bibinfo  {journal}
  {Naturwissenschaften}\ }\textbf {\bibinfo {volume} {13}},\ \bibinfo {pages}
  {627} (\bibinfo {year} {1925})}\BibitemShut {NoStop}%
\bibitem [{\citenamefont {Reiche}\ and\ \citenamefont
  {Thomas}(1925)}]{reich25.01}%
  \BibitemOpen
  \bibfield  {author} {\bibinfo {author} {\bibfnamefont {F.}~\bibnamefont
  {Reiche}}\ and\ \bibinfo {author} {\bibfnamefont {u.~W.}\ \bibnamefont
  {Thomas}},\ }\href@noop {} {\bibfield  {journal} {\bibinfo  {journal} {Z.
  Physik}\ }\textbf {\bibinfo {volume} {34}},\ \bibinfo {pages} {879} (\bibinfo
  {year} {1925})}\BibitemShut {NoStop}%
\bibitem [{\citenamefont {Kuhn}(1925)}]{kuhn25.01}%
  \BibitemOpen
  \bibfield  {author} {\bibinfo {author} {\bibfnamefont {W.}~\bibnamefont
  {Kuhn}},\ }\href@noop {} {\bibfield  {journal} {\bibinfo  {journal}
  {Zeitschrift fur Physik A: Hadrons and Nuclei}\ }\textbf {\bibinfo {volume}
  {33}},\ \bibinfo {pages} {408} (\bibinfo {year} {1925})}\BibitemShut
  {NoStop}%
\bibitem [{\citenamefont {Kuzyk}(2003{\natexlab{b}})}]{kuzyk03.02}%
  \BibitemOpen
  \bibfield  {author} {\bibinfo {author} {\bibfnamefont {M.~G.}\ \bibnamefont
  {Kuzyk}},\ }\href@noop {} {\bibfield  {journal} {\bibinfo  {journal} {Phys.
  Rev. Lett.}\ }\textbf {\bibinfo {volume} {90}},\ \bibinfo {pages} {039902}
  (\bibinfo {year} {2003}{\natexlab{b}})}\BibitemShut {NoStop}%
\bibitem [{\citenamefont {Kuzyk}(2000{\natexlab{b}})}]{kuzyk00.02}%
  \BibitemOpen
  \bibfield  {author} {\bibinfo {author} {\bibfnamefont {M.~G.}\ \bibnamefont
  {Kuzyk}},\ }\href@noop {} {\bibfield  {journal} {\bibinfo  {journal} {Opt.
  Lett.}\ }\textbf {\bibinfo {volume} {25}},\ \bibinfo {pages} {1183} (\bibinfo
  {year} {2000}{\natexlab{b}})}\BibitemShut {NoStop}%
\bibitem [{\citenamefont {Clays}(2001)}]{clays01.01}%
  \BibitemOpen
  \bibfield  {author} {\bibinfo {author} {\bibfnamefont {K.}~\bibnamefont
  {Clays}},\ }\href@noop {} {\bibfield  {journal} {\bibinfo  {journal} {Opt.
  Lett.}\ }\textbf {\bibinfo {volume} {26}},\ \bibinfo {pages} {1699} (\bibinfo
  {year} {2001})}\BibitemShut {NoStop}%
\bibitem [{\citenamefont {Kuzyk}\ \emph {et~al.}(2013)\citenamefont {Kuzyk},
  \citenamefont {Perez-Moreno},\ and\ \citenamefont {Shafei}}]{kuzyk13.01}%
  \BibitemOpen
  \bibfield  {author} {\bibinfo {author} {\bibfnamefont {M.~G.}\ \bibnamefont
  {Kuzyk}}, \bibinfo {author} {\bibfnamefont {J.}~\bibnamefont {Perez-Moreno}},
  \ and\ \bibinfo {author} {\bibfnamefont {S.}~\bibnamefont {Shafei}},\
  }\href@noop {} {\bibfield  {journal} {\bibinfo  {journal} {Phys. Rep}\
  }\textbf {\bibinfo {volume} {529}},\ \bibinfo {pages} {297} (\bibinfo {year}
  {2013})}\BibitemShut {NoStop}%
\bibitem [{\citenamefont {Kuzyk}(2006)}]{kuzyk06.03}%
  \BibitemOpen
  \bibfield  {author} {\bibinfo {author} {\bibfnamefont {M.~G.}\ \bibnamefont
  {Kuzyk}},\ }\href@noop {} {\bibfield  {journal} {\bibinfo  {journal} {J. Chem
  Phys.}\ }\textbf {\bibinfo {volume} {125}},\ \bibinfo {pages} {154108}
  (\bibinfo {year} {2006})}\BibitemShut {NoStop}%
\bibitem [{\citenamefont {{Lytel}}(2016)}]{lytel16.02}%
  \BibitemOpen
  \bibfield  {author} {\bibinfo {author} {\bibfnamefont {R.}~\bibnamefont
  {{Lytel}}},\ }\href@noop {} {\bibfield  {journal} {\bibinfo  {journal} {J.
  Opt. Soc. Am. B}\ }\textbf {\bibinfo {volume} {33}},\ \bibinfo {pages} {E66}
  (\bibinfo {year} {2016})}\BibitemShut {NoStop}%
\bibitem [{\citenamefont {Watkins}\ and\ \citenamefont
  {Kuzyk}(2009)}]{watki09.01}%
  \BibitemOpen
  \bibfield  {author} {\bibinfo {author} {\bibfnamefont {D.~S.}\ \bibnamefont
  {Watkins}}\ and\ \bibinfo {author} {\bibfnamefont {M.~G.}\ \bibnamefont
  {Kuzyk}},\ }\href@noop {} {\bibfield  {journal} {\bibinfo  {journal} {J.
  Chem. Phys.}\ }\textbf {\bibinfo {volume} {131}},\ \bibinfo {pages} {064110}
  (\bibinfo {year} {2009})}\BibitemShut {NoStop}%
\bibitem [{\citenamefont {Watkins}\ and\ \citenamefont
  {Kuzyk}(2011)}]{watki11.01}%
  \BibitemOpen
  \bibfield  {author} {\bibinfo {author} {\bibfnamefont {D.~S.}\ \bibnamefont
  {Watkins}}\ and\ \bibinfo {author} {\bibfnamefont {M.~G.}\ \bibnamefont
  {Kuzyk}},\ }\href@noop {} {\bibfield  {journal} {\bibinfo  {journal} {J.
  Chem. Phys.}\ }\textbf {\bibinfo {volume} {134}},\ \bibinfo {pages} {094109}
  (\bibinfo {year} {2011})}\BibitemShut {NoStop}%
\bibitem [{\citenamefont {Atherton}\ \emph {et~al.}(2012)\citenamefont
  {Atherton}, \citenamefont {Lesnefsky}, \citenamefont {Wiggers},\ and\
  \citenamefont {Petschek}}]{ather12.01}%
  \BibitemOpen
  \bibfield  {author} {\bibinfo {author} {\bibfnamefont {T.}~\bibnamefont
  {Atherton}}, \bibinfo {author} {\bibfnamefont {J.}~\bibnamefont {Lesnefsky}},
  \bibinfo {author} {\bibfnamefont {G.}~\bibnamefont {Wiggers}}, \ and\
  \bibinfo {author} {\bibfnamefont {R.}~\bibnamefont {Petschek}},\ }\href@noop
  {} {\bibfield  {journal} {\bibinfo  {journal} {J. Opt. Soc. Am. B}\ }\textbf
  {\bibinfo {volume} {29}},\ \bibinfo {pages} {513} (\bibinfo {year}
  {2012})}\BibitemShut {NoStop}%
\bibitem [{\citenamefont {Burke}\ \emph {et~al.}(2013)\citenamefont {Burke},
  \citenamefont {Atherton}, \citenamefont {Lesnefsky},\ and\ \citenamefont
  {Petschek}}]{burke13.01}%
  \BibitemOpen
  \bibfield  {author} {\bibinfo {author} {\bibfnamefont {C.~J.}\ \bibnamefont
  {Burke}}, \bibinfo {author} {\bibfnamefont {T.~J.}\ \bibnamefont {Atherton}},
  \bibinfo {author} {\bibfnamefont {J.}~\bibnamefont {Lesnefsky}}, \ and\
  \bibinfo {author} {\bibfnamefont {R.~G.}\ \bibnamefont {Petschek}},\ }\href
  {\doibase 10.1364/JOSAB.30.001438} {\bibfield  {journal} {\bibinfo  {journal}
  {J. Opt. Soc. Am. B}\ }\textbf {\bibinfo {volume} {30}},\ \bibinfo {pages}
  {1438} (\bibinfo {year} {2013})}\BibitemShut {NoStop}%
\bibitem [{\citenamefont {Lytel}\ and\ \citenamefont
  {Kuzyk}(2013)}]{lytel13.04}%
  \BibitemOpen
  \bibfield  {author} {\bibinfo {author} {\bibfnamefont {R.}~\bibnamefont
  {Lytel}}\ and\ \bibinfo {author} {\bibfnamefont {M.~G.}\ \bibnamefont
  {Kuzyk}},\ }\href@noop {} {\bibfield  {journal} {\bibinfo  {journal} {Journal
  of Nonlinear Optical Physics \& Materials}\ }\textbf {\bibinfo {volume} {22}}
  (\bibinfo {year} {2013})}\BibitemShut {NoStop}%
\bibitem [{\citenamefont {Kuzyk}\ and\ \citenamefont
  {Kuzyk}(2008)}]{kuzyk08.01}%
  \BibitemOpen
  \bibfield  {author} {\bibinfo {author} {\bibfnamefont {M.~C.}\ \bibnamefont
  {Kuzyk}}\ and\ \bibinfo {author} {\bibfnamefont {M.~G.}\ \bibnamefont
  {Kuzyk}},\ }\href@noop {} {\bibfield  {journal} {\bibinfo  {journal} {J. Opt.
  Soc. Am. B.}\ }\textbf {\bibinfo {volume} {25}},\ \bibinfo {pages} {103}
  (\bibinfo {year} {2008})}\BibitemShut {NoStop}%
\bibitem [{\citenamefont {Kuzyk}(2005)}]{kuzyk05.02}%
  \BibitemOpen
  \bibfield  {author} {\bibinfo {author} {\bibfnamefont {M.~G.}\ \bibnamefont
  {Kuzyk}},\ }\href@noop {} {\bibfield  {journal} {\bibinfo  {journal} {Phys.
  Rev. A}\ }\textbf {\bibinfo {volume} {72}},\ \bibinfo {pages} {053819}
  (\bibinfo {year} {2005})}\BibitemShut {NoStop}%
\bibitem [{\citenamefont {Champagne}\ and\ \citenamefont
  {Kirtman}(2005)}]{champ05.01}%
  \BibitemOpen
  \bibfield  {author} {\bibinfo {author} {\bibfnamefont {B.}~\bibnamefont
  {Champagne}}\ and\ \bibinfo {author} {\bibfnamefont {B.}~\bibnamefont
  {Kirtman}},\ }\href@noop {} {\bibfield  {journal} {\bibinfo  {journal} {Phys.
  Rev. Lett.}\ }\textbf {\bibinfo {volume} {95}},\ \bibinfo {pages} {109401}
  (\bibinfo {year} {2005})}\BibitemShut {NoStop}%
\bibitem [{\citenamefont {Champagne}\ and\ \citenamefont
  {Kirtman}(2006)}]{champ06.01}%
  \BibitemOpen
  \bibfield  {author} {\bibinfo {author} {\bibfnamefont {B.}~\bibnamefont
  {Champagne}}\ and\ \bibinfo {author} {\bibfnamefont {B.}~\bibnamefont
  {Kirtman}},\ }\href@noop {} {\bibfield  {journal} {\bibinfo  {journal} {J .
  Chem. Phys.}\ }\textbf {\bibinfo {volume} {125}},\ \bibinfo {pages} {024101}
  (\bibinfo {year} {2006})}\BibitemShut {NoStop}%
\bibitem [{\citenamefont {Shafei}\ and\ \citenamefont
  {Kuzyk}(2011)}]{shafe11.01}%
  \BibitemOpen
  \bibfield  {author} {\bibinfo {author} {\bibfnamefont {S.}~\bibnamefont
  {Shafei}}\ and\ \bibinfo {author} {\bibfnamefont {M.~G.}\ \bibnamefont
  {Kuzyk}},\ }\href@noop {} {\bibfield  {journal} {\bibinfo  {journal} {J. Opt.
  Soc. Am. B}\ }\textbf {\bibinfo {volume} {28}},\ \bibinfo {pages} {882}
  (\bibinfo {year} {2011})}\BibitemShut {NoStop}%
\bibitem [{\citenamefont {Dawson}\ and\ \citenamefont
  {Kuzyk}(2016)}]{dawso16.02}%
  \BibitemOpen
  \bibfield  {author} {\bibinfo {author} {\bibfnamefont {N.~J.}\ \bibnamefont
  {Dawson}}\ and\ \bibinfo {author} {\bibfnamefont {M.~G.}\ \bibnamefont
  {Kuzyk}},\ }\href@noop {} {\bibfield  {journal} {\bibinfo  {journal} {J. Opt.
  Soc. Am. B}\ }\textbf {\bibinfo {volume} {33}},\ \bibinfo {pages} {E83}
  (\bibinfo {year} {2016})}\BibitemShut {NoStop}%
\bibitem [{\citenamefont {{Lytel}}\ \emph {et~al.}(2015)\citenamefont
  {{Lytel}}, \citenamefont {{Mossman}},\ and\ \citenamefont
  {{Kuzyk}}}]{lytel15.02}%
  \BibitemOpen
  \bibfield  {author} {\bibinfo {author} {\bibfnamefont {R.}~\bibnamefont
  {{Lytel}}}, \bibinfo {author} {\bibfnamefont {S.}~\bibnamefont {{Mossman}}},
  \ and\ \bibinfo {author} {\bibfnamefont {M.}~\bibnamefont {{Kuzyk}}},\
  }\href@noop {} {\bibfield  {journal} {\bibinfo  {journal} {Optics Letters}\
  }\textbf {\bibinfo {volume} {40}},\ \bibinfo {pages} {4735} (\bibinfo {year}
  {2015})},\ \Eprint {http://arxiv.org/abs/1508.06560} {arXiv:1508.06560
  [physics.optics]} \BibitemShut {NoStop}%
\bibitem [{\citenamefont {{\v{S}}tefko}\ \emph {et~al.}(2013)\citenamefont
  {{\v{S}}tefko}, \citenamefont {Tzirakis}, \citenamefont {Breiten},
  \citenamefont {Ebert}, \citenamefont {Dumele}, \citenamefont {Schweizer},
  \citenamefont {Gisselbrecht}, \citenamefont {Boudon}, \citenamefont {Beels},
  \citenamefont {Biaggio},\ and\ \citenamefont {Diederich}}]{stefk13.01}%
  \BibitemOpen
  \bibfield  {author} {\bibinfo {author} {\bibfnamefont {M.}~\bibnamefont
  {{\v{S}}tefko}}, \bibinfo {author} {\bibfnamefont {M.~D.}\ \bibnamefont
  {Tzirakis}}, \bibinfo {author} {\bibfnamefont {B.}~\bibnamefont {Breiten}},
  \bibinfo {author} {\bibfnamefont {M.-O.}\ \bibnamefont {Ebert}}, \bibinfo
  {author} {\bibfnamefont {O.}~\bibnamefont {Dumele}}, \bibinfo {author}
  {\bibfnamefont {W.~B.}\ \bibnamefont {Schweizer}}, \bibinfo {author}
  {\bibfnamefont {J.-P.}\ \bibnamefont {Gisselbrecht}}, \bibinfo {author}
  {\bibfnamefont {C.}~\bibnamefont {Boudon}}, \bibinfo {author} {\bibfnamefont
  {M.~T.}\ \bibnamefont {Beels}}, \bibinfo {author} {\bibfnamefont
  {I.}~\bibnamefont {Biaggio}}, \ and\ \bibinfo {author} {\bibfnamefont
  {F.}~\bibnamefont {Diederich}},\ }\href@noop {} {\bibfield  {journal}
  {\bibinfo  {journal} {Chemistry-A European Journal}\ }\textbf {\bibinfo
  {volume} {19}},\ \bibinfo {pages} {12693} (\bibinfo {year}
  {2013})}\BibitemShut {NoStop}%
\bibitem [{\citenamefont {Sullivan}\ \emph {et~al.}(2016)\citenamefont
  {Sullivan}, \citenamefont {Mossman},\ and\ \citenamefont
  {Kuzyk}}]{sulli16.02}%
  \BibitemOpen
  \bibfield  {author} {\bibinfo {author} {\bibfnamefont {D.}~\bibnamefont
  {Sullivan}}, \bibinfo {author} {\bibfnamefont {S.}~\bibnamefont {Mossman}}, \
  and\ \bibinfo {author} {\bibfnamefont {M.~G.}\ \bibnamefont {Kuzyk}},\
  }\href@noop {} {\bibfield  {journal} {\bibinfo  {journal} {J. Opt. Soc. Am.
  B}\ }\textbf {\bibinfo {volume} {33}},\ \bibinfo {pages} {E143} (\bibinfo
  {year} {2016})}\BibitemShut {NoStop}%
\end{thebibliography}%
%


\end{document}